\documentclass{emulateapj}

\usepackage{graphicx}
\usepackage{epstopdf, epsfig}
\usepackage{natbib}
\usepackage{multirow}
\usepackage{amsmath,amssymb}
\usepackage{rotating}

\begin{document}
\submitted{Accepted to The Astrophysical Journal}
\journalinfo{Accepted to The Astrophysical Journal}

\title{SMA observations on faint Submillimeter Galaxies with $S_{850} < 2$\,mJy: Ultra Dusty Low-Luminosity Galaxies at High Redshift}
\author{Chian-Chou Chen\altaffilmark{1,2}, Lennox L. Cowie\altaffilmark{1}, Amy J. Barger\altaffilmark{1,3,4},  Wei-Hao Wang\altaffilmark{5}, Jonathan P. Williams\altaffilmark{1}}
\altaffiltext{1}{Institute for Astronomy, University of Hawaii, 2680 Woodlawn Drive, Honolulu, HI 96822.}
\altaffiltext{2}{Institute for Computational Cosmology, Department of Physics, Durham University, South Road, Durham DH1 3LE, UK}
\altaffiltext{3}{Department of Astronomy, University of Wisconsin-Madison, 475 North Charter Street, Madison, WI 53706.}
\altaffiltext{4}{Department of Physics and Astronomy, University of Hawaii, 2505 Correa Road, Honolulu, HI 96822.}
\altaffiltext{5}{Academia Sinica Institute of Astronomy and Astrophysics, P.O. Box 23-141, Taipei 10617, Taiwan.}
\subjectheadings{cosmology: observations|  galaxies: formation  |  galaxies: starburst  |  gravitational lensing: strong | submillimeter: galaxies }

\begin{abstract}
We obtained SMA observations of eight faint (intrinsic 850\,$\mu$m fluxes $<$ 2\,mJy) submillimeter galaxies (SMGs) discovered in SCUBA images of the massive lensing cluster fields A370, A2390, and A1689 and detected five. In total, we obtain 5 SMA detections, all of which have de-lensed fluxes $<$1\,mJy with estimated total infrared luminosities 10$^{10}-10^{12}$\,$L_\odot$, comparable to luminous infrared galaxies (LIRGs) and normal star-forming galaxies. Based on the latest number counts, these galaxies contribute $\sim$70\% of the 850\,$\mu$m extragalactic background light and represent the dominant star-forming galaxy population in the dusty universe. 
However, only {40}$^{+30}_{-16}$\% of our faint SMGs would be detected in deep optical or near-infrared surveys, which suggests many of these sources are at high redshifts ($z \gtrsim 3$) or extremely dusty, and they are not included in current star formation history estimates.
\end{abstract}

\section{Introduction}
{Bright submillimeter galaxies (SMGs; $S_{850\,\mu{\rm m}}>2$~mJy) have luminosities corresponding to those of  local ultraluminous infrared galaxies (ULIRGs; $L_{IR}>10^{12}~L_\odot$). Many of these sources have very faint 
optical or near-infrared counterparts, reflecting their large dust contents and high redshifts (e.g., \citealt{Younger:2009p9502, Barger:2012lr, Walter:2012qy,Simpson:2013lr}). However, as we move to lower 
submillimeter fluxes, we might expect the SMG population to {become less dusty, because the spectral energy distributions become more UV dominated for lower infrared luminosities (e.g., \citealt{Chary:2001aa}), and hence for the faint SMG population to be substantially overlapped with the optically selected population.} The goal of the present paper is to test this expectation, given how critical it is to {obtain an} accurate determination of the star formation history.

Many blank-field 850\,$\mu$m surveys have been made with ground-based, single-dish telescopes, such as the James Clerk Maxwell Telescope (JCMT) and the Atacama Pathfinder Experiment (APEX)
\citep{Barger:1998p13566,Hughes:1998p9666,Eales:1999p9715,Eales:2000uq, Scott:2002p6539, Smail:2002p6793, Borys:2003p6612, Serjeant:2003lr, Webb:2003p6591, Wang:2004p2270, Coppin:2006p9123, Weis:2009qy, Casey:2013aa, Geach:2013kx,Barger:2014aa}. 
Detailed follow-up continuum and emission line studies have shown that many of these SMGs 
(review by \citealt{Blain:2002p8120}) are at high redshifts with $2 < z < 5$ (e.g., \citealt{Chapman:2005p5778, Wardlow:2011qy, Walter:2012qy, Barger:2012lr}), gas rich ($M_{gas} > 10^{10}~M_\odot$; e.g., \citealt{Greve:2005p6788, Bothwell:2013lp}), highly clustered (e.g., \citealt{Scott:2006lr, Hickox:2012kk}), have both disk-like and merger-like morphologies (e.g., \citealt{Tacconi:2008p9334, Hodge:2012fk}), and may be the progenitors of massive elliptical galaxies (e.g., \citealt{Lilly:1999lr, Fu:2013mm, Simpson:2013lr}). Using stellar population synthesis models, the derived stellar masses of SMGs are typically in the range $\sim$\,10$^{11}$--10$^{12}\,M_\odot$ (e.g., \citealt{Borys:2005rt, Dye:2008ys, Michaowski:2012fr}). Incorporating the shorter wavelength data from the {\em Herschel Space Observatory\/} (hereafter, {\it Herschel}), \citet{Magnelli:2012lr} argued that 850\,$\mu$m selected SMGs are diverse in dust temperature (20--60\,K). However, \citet{Barger:2014aa}
using a small, uniformly selected sample of 850~$\mu$m sources lying in the flux range 3--15~mJy, found a much smaller range of temperatures. X-ray observations of 850\,$\mu$m selected SMGs with radio and/or mid-infrared (MIR) counterparts have revealed, on average, order of magnitude lower X-ray-to-far-infrared (FIR) luminosity ratios for SMGs than for AGN dominated quasars, and FIR luminosity outputs dominated by star formation (e.g., \citealt{Alexander:2005p6453, Laird:2010lr, Symeonidis:2011no, Wang:2013aa}). Despite their rareness, the extreme star formation rates (SFR $\sim 500-10000~M_\odot$~yr$^{-1}$) of SMGs make them substantial contributors to star formation in the early universe (e.g., \citealt{Barger:2000p2144, Barger:2012lr,Barger:2014aa, Chapman:2005p5778, Wang:2006p2031, Serjeant:2008p6307, Wardlow:2011qy,  Casey:2013aa, Swinbank:2013ul}). 

\begin{table*}[t]
 \begin{center}
 \caption{SMA Observations}
\resizebox{18cm}{!} {
\begin{tabular}{cllrrrlll}
\hline
\hline
 I.D.  &  Source   Name         &  Track Dates                         &       Beam$^{c}$   &   Beam$^{c}$   &  $\sigma^{c}$ & Flux              &  Passband       &  Gain                    \\
       &                        &                                &       FWHM  &   P.A.  &   {(mJy/ }    &  Calibrator(s)     &  Calibrator(s)  &  Calibrator(s)           \\
       &                        &                                &    ($'' \times ''$)  &  (deg)  &    {beam)}  &                    &                 &                          \\
 (1) &	(2)	    &			(3)	& (4)&(5)&{(6)}&(7)&(8)&(9) \\
\hline
 {Chen-1}  &  \#4$^a$                &  20121030, 20121109, 20121111  &  2.08$\times$2.06  &   55.8  &           {0.61} & Neptune           &  bllac          &  0309+104, 0339-017      \\
 {Chen-2}  &  \#12$^{a,g}$              &  20090626$^{d}$, 20090627$^{d}$, 20091016  &  1.98$\times$1.63  &   44.8  &    {0.80} &       Callisto, Uranus  &  3c454.3        &  3c454.3, 2148+069       \\
 {Chen-3}  &  \#14$^a$               &  20110524, 20110925            &  2.22$\times$1.80  &  -33.3  &          {1.01} & Uranus            &  3c454.3, 3c84  &  3c454.3, 2203+174       \\
 {Chen-4}  &  SMM J131128.6-012036$^{b}$   &  20120506, 20120509            &  1.93$\times$1.74  &   80.7  &         {0.92} & Neptune, Titan    &  bllac, 3c279   &  3c279, 3c273, 1337-129  \\
 {Chen-5}  &  SMM J131129.1-012049$^{b}$   &  20120508                      &  2.05$\times$1.78  &   78.6  &          {1.10} & Titan             &  bllac          &  3c273, 1337-129         \\
 {Chen-6}  &  SMM J131132.0-011955$^{b}$   &  20130226, 20130303            &  2.36$\times$1.86  &  -43.7  &         {0.74} & Titan, Callisto   &  bllac, 3c84    &  3c273, 1337-129         \\
 {Chen-7} &  SMM J131134.1-012021$^{b,}$$^{e}$   &  20130226, 20130303, 20130306  &  2.36$\times$1.98  &  -43.4  &         {0.65} & Titan, Callisto   &  bllac, 3c84    &  3c273, 1337-129         \\
 {Chen-8}  &  SMM J131135.1-012018$^{b,}$$^{e}$   &  20130226, 20130303, 20130306  &  2.36$\times$1.98  &  -43.4  &         {1.00$^f$} & Titan, Callisto   &  bllac, 3c84    &  3c273, 1337-129         \\
 \hline
 &&&&&&&&\\
 \multicolumn{8}{l}{$^{a}$ Sources first identified by \citet{Cowie:2002p2075} with SCUBA.}\\
  \multicolumn{8}{l}{$^{b}$ Sources first identified by \citet{Knudsen:2008p3824} with SCUBA.}\\
  \multicolumn{8}{l}{$^{c}$ Results from all the tracks combined on a given source. {The values of $\sigma_{t}$ are the theoretical sensitivities, while the values of $\sigma$ are the r.m.s.} }\\
  \multicolumn{8}{l}{{of the dirty maps.}} \\
  \multicolumn{8}{l}{$^{d}$ The bandwidth per each sideband was 2\,GHz on these tracks, whereas it was 4\,GHz for the others.}\\
  \multicolumn{9}{l}{$^{e}$ Both Chen-7 and  Chen-8 were observed in the same dataset. The field of view of the SMA primary beam covers these two sources.}\\
  \multicolumn{8}{l}{$^{f}$ The sensitivity at the SCUBA position shown in Table \ref{src}.}\\
  \multicolumn{8}{l}{{$^{g}$ This source was named A2390-3 in \citet{Chen:2011p11605}.}}\\
 \end{tabular}
 }
 \label{sma}
 \end{center}
\end{table*}

However, the blank-field SMGs 
only contribute $20-30$\% of the 850~$\mu$m extragalactic background light (EBL; e.g., \citealt{Barger:1999p6485, Coppin:2006p9123, Chen:2013fk, Chen:2013gq}), which is the integrated emission from all extragalactic sources along the line-of-sight. Thus, the bulk of dusty star formation is still unresolved, and determining the characteristics of the faint SMGs with typical $L_{IR} < 10^{12}~L_\odot$ that emit most of the 850~$\mu$m EBL is needed for a full understanding of the cosmic star formation history. 

Unfortunately, the poor resolution at 850~$\mu$m (e.g., $\sim14''$ FWHM on the 15~m JCMT) prevents us from directly measuring the faint SMGs below the 2~mJy confusion limit in blank fields. Almost all of our knowledge about faint SMGs comes from ground-based observations with single-dish telescopes in the fields of massive lensing clusters (e.g., \citealt{Smail:1997p6820, Cowie:2002p2075, Kneib:2004uq, Knudsen:2009p9556, Knudsen:2010lr, Boone:2013fr, Chen:2013fk}). Due to the presence of the intervening cluster mass, the intrinsically faint fluxes of background sources are gravitationally amplified to a detectable level, and the confusion limit is reduced by the expansion of the source plane. Faint SMGs with fluxes between 0.1 and 2~mJy have been detected in this way. Although the number of faint SMGs that have been discovered in lensing fields is small compared to the number of bright SMGs that have been found in blank fields, their number density indicates that they contribute $\sim$ 70\% of the 850~$\mu$m EBL \citep{Blain:1999p7279,Cowie:2002p2075, Knudsen:2008p3824, Zemcov:2010uq, Chen:2013fk, Chen:2013gq}.

\begin{table*}[t]
\begin{center}
 \caption{Spitzer Super Mosaics}
\begin{tabular}{lrrrrrrrrrr}
 \hline
\hline
\multirow{2}{*}{ Field}  &   3.6\,$\mu$m      &  Sensitivity  	&   4.5\,$\mu$m  		&  Sensitivity  	&   5.8\,$\mu$m  	&  Sensitivity  	&   8.0\,$\mu$m  	&  Sensitivity  	&   24\,$\mu$m  	&  Sensitivity  \\
				   & (ks)	  &($\mu$Jy)	& (ks)		&($\mu$Jy)		& (ks)	&($\mu$Jy)		& (ks)	&($\mu$Jy)		& (ks)	&($\mu$Jy)	 \\
\hline
 A370   &  13.0  &          0.3  &  12.3  &          0.3  &  13.1  &          1.5  &  11.8  &          1.5  &  2.6  &           31  \\
 A2390  &   6.3  &          0.4  &   6.9  &          0.4  &   6.3  &          1.7  &   6.6  &          2.5  &  0.6  &           41  \\
 A1689  &  10.8  &          0.3  &  10.7  &          0.3  &  10.7  &          1.4  &  10.3  &          1.5  &  0.5  &           36  \\
\hline
 &&&&&&&&&&\\
 \multicolumn{11}{l}{Notes: The exposures are the median exposure time in each super mosaic in kiloseconds (ks). The sensitivities represent}\\
  \multicolumn{11}{l}{\hspace{9mm} 1\,$\sigma$ errors of the aperture photometry.}\\
\end{tabular}
 \label{spitzer}
\end{center}
\end{table*}

While many faint SMGs will eventually be observed with extremely sensitive submillimeter interferometric arrays, such as the Atacama Large Millimeter/submillimeter Array (ALMA), we can already begin to investigate some fundamental questions about faint SMGs using sources discovered in lensing fields. For example, using high spatial resolution submillimeter continuum observations with the Submillimeter Array (SMA), we can pin down the exact location of faint SMGs discovered with the SCUBA instrument on the JCMT \citep{Holland:1999fk} and find their true counterparts, if any, at other wavelengths. Once we know the correct counterparts, then we can study the properties of the faint SMGs, such as their colors and redshift distribution. Perhaps even more exciting, we can estimate the fraction of faint SMGs that are completely
hidden from current optical/near-infrared (NIR) observations.

In this paper, we use SMA observations of an unbiased sample of eight highly amplified and intrinsically faint SCUBA-detected SMGs discovered in the fields of three massive lensing clusters, Abell 370 (A370), Abell 2390 (A2390), and Abell 1689 (A1689), to study the faint SMG population. The SCUBA sources are taken from the catalogs of \citet{Cowie:2002p2075} and \citet{Knudsen:2008p3824}. 
{While previous studies focused on a few individual sources that had optical/NIR counterparts (e.g., \citealt{Kneib:2004uq, Knudsen:2010lr}), our only specifications are discovery in single-dish surveys and high amplifications ({$> $}\,3), meaning intrinsic 850\,$\mu$m fluxes expected to be less than 2\,mJy. } 

In \citet{Chen:2011p11605}, we already presented the SMA observations of one of the SCUBA sources in our sample, {Chen-2}. There we showed that despite the identification of likely candidate counterparts using a traditional $p-$value analysis, once we had the accurate source position from the SMA, we could see that there were no viable counterparts from the optical to the radio. The lack of a deep radio counterpart led us to conclude that {Chen-2} could be at a very high redshift ($z>4$). This surprising result suggests that, while the NIR stacking analyses show that a large percentage ($\sim$ 50\%) of the 850\,$\mu$m EBL could come from sources at $z<1.5$ \citep{Wang:2006p2031,Serjeant:2008p6307}, a number of faint SMGs ($<$ 2\,mJy) may lie at high redshifts, and they are likely missed by current optical/NIR observations. However, this suggestion is based on one source, and the results {may not be representative of our selected sample as a whole}. In this paper, we report on the full results of our analysis. 

We describe the SMA data and data reduction in Section~2. We give our results in Section~3. In Section~4, we discuss the implications of our results, and in Section~5, we summarize the paper. We assume the Wilkinson Microwave Anisotropy Probe cosmology throughout: H$_0$ = 70.5 km s$^{-1}$ Mpc$^{-1}$, $\Omega_M$ = 0.27, and $\Omega_\Lambda$ = 0.73 \citep{Larson:2011ys}.

\section{Observations and Data Reduction}
\subsection{SMA Observations}
We conducted SMA\footnote[6]{The Submillimeter Array is a joint project between the Smithsonian Astrophysical Observatory and the Academia Sinica Institute of Astronomy and Astrophysics and is funded by the Smithsonian Institution and the Academia Sinica.} \citep{Ho:2004p8376} observations in compact configuration (16--77 m baselines) of a sample of 8 highly-amplified SCUBA sources in the massive lensing cluster fields A370, A2390, and A1689. We expected the intrinsic 850\,$\mu$m fluxes of the SMGs to be lower than the confusion limit ($\sim$2\,mJy) based on their discovered locations and the lensing models. We tuned our observations to the low spectral resolution mode (32 frequency channels per chunk) with local oscillator frequency at 343\,GHz, so we label our observations as 870\,$\mu$m.

We summarize our data in Table \ref{sma}. In Column~1, we give the source {IDs, which are used in the rest of the paper}; in Column~2, the SCUBA source name from \citet{Cowie:2002p2075} or \citet{Knudsen:2008p3824}; in Column~3, the dates when the SMA tracks were observed; in Column~4, the beam size; in Column~5, the beam position angle; {in Column\, {6}, the r.m.s. value of the dirty map at the phase center ($\sigma$)}; in Column~{{7}, the flux calibrator(s); in Column~{8}, the passband calibrator(s); and in Column~{9}, the gain calibrators. 

We used the data reduction package MIR to calibrate the visibilities. The visibility data were first weighted by the system temperatures ($T_{sys}$). Then the bandpass responses were measured and corrected through observations of bright quasars (Column~{8}). The phase changes were monitored using neighboring known point sources (Column~{9}). Given that all observations were conducted under good weather conditions with Precipitable Water Vapor (PWV) of $\le$ 1\,mm, the phases are stable, and the phase calibrations using multiple (mostly two) calibrators agree with each other. 
We used planets for the flux calibration. The typical uncertainty of the flux calibration is $\sim$\,10\%. 

\begin{figure}
 \begin{center}
    \leavevmode
      \includegraphics[scale=0.56]{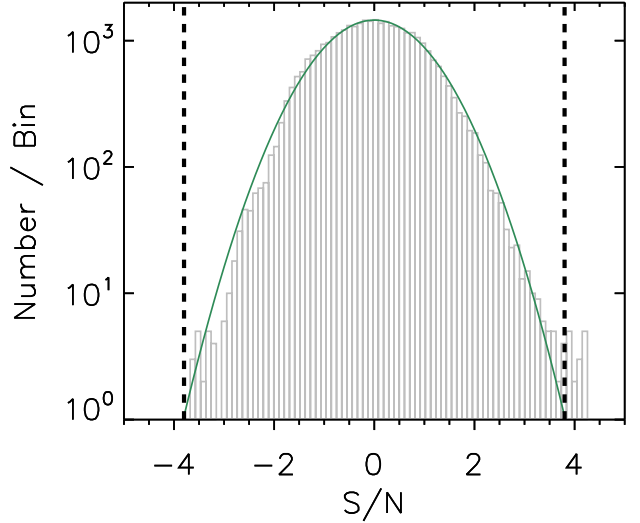}
       \caption{S/N histogram of Chen-3 for the pixels located within the SMA primary beam. The S/N map is generated by dividing the dirty maps by the r.m.s. values of the dirty maps. The green curve shows the ideal Gaussian distribution having r.m.s. of 1, {and the dashed vertical lines mark $\pm$ 3.8\,$\sigma$. This figure shows that the detection threshold of S/N $>$ 3.8 used in this paper is robust.}}
    \label{histo}
  \end{center}
\end{figure}

We used the calibrated visibilities to produce the images through the MIRIAD routines \citep{Sault:1995p6974}. We combined the visibilities from all the available tracks for each source. 
We used the routine INVERT with natural weighting on the baselines to perform the inverse Fourier transformations on the visibilities in order to produce the dirty maps and the synthesized dirty beam images with 0$\farcs$2 grids. The natural weighting scheme provides the best signal-to-noise ratio (S/N) at the cost of no sidelobe suppression and slightly poorer resolution. The typical FWHM of our synthesized beam is $\sim2'''$. We also performed multi-frequency synthesis during the inverse Fourier transformations, which gives better coverage in the frequency-dependent uv coordinate.

\begin{table*}
\begin{center}
 \caption{The properties of the SMA observed sources}
\begin{tabular}{cccccccc}
\hline
\hline
 {I.D.}         &  SMA R.A.  &  SMA Decl.  &  SCUBA R.A.  &  SCUBA Decl.  &  SCUBA  &  SCUBA-2  &  SMA   \\
                       &  (J2000)   &  (J2000)    &  (J2000)     &  (J2000)      &  850\,$\mu$m    &  850\,$\mu$m  & 870\,$\mu$m   \\
                       &  (h m s)   &  (d m s)    &  (h m s)     &  (d m s)      &  (mJy)  &  (mJy)   &  (mJy)  \\
\hline
 {Chen-1}                &    $\cdots$        &    $\cdots$         &     02 39 53.83         &     -01 33 37.0        &   2.17$\pm$0.57      &  -0.01$\pm$0.51        &  $<${2.52 (4\,${\sigma}$)}      \\
 {Chen-2}               &   21 53 35.16         &   17 41 06.1          &  21 53 35.48            &  17 41 09.3             &   3.24$\pm$0.78      &   {2.35$\pm$0.51}       & 3.96$\pm${1.01}       \\
 {Chen-3}               &   21 53 34.37         &   17 42 01.5          &    21 53 34.15          &   17 42 02.3            &    2.64$\pm$0.72     &   {1.93$\pm$0.51}       &   4.72$\pm${0.89}      \\
 {Chen-4}  &    $\cdots$        &   $\cdots$          &   13 11 28.6         &   -01 20 36            &  2.6$\pm$0.8       &   -0.95$\pm$0.48       &  $<${3.72 (4\,${\sigma}$)}       \\
 {Chen-5}  &   13 11 29.22         &   --01 20 44.5          &  13 11 29.1          &   -01 20 49            &  4.7$\pm$0.8       &   4.39$\pm$0.48       &   5.25$\pm${0.87}      \\
 {Chen-6}  &13 11 31.93   & --01 19 55.1 &  13 11 32.0          &    -01 19 55          &   3.3$\pm$1.0      &    3.28$\pm$0.50      & 2.73$\pm${0.81}\\
{Chen-7}  &   $\cdots$         &   $\cdots$         &  13 11 34.1          &    -01 20 21           &  3.2$\pm$1.0       &   4.32$\pm$0.52       &   $<${2.56 (4\,${\sigma}$)}      \\
{Chen-8}   & 13 11 34.95         &  --01 20 17.2   &  13 11 35.1          &    -01 20 18           &  4.9$\pm$1.6       &   4.15$\pm$0.54       &   3.92$\pm${0.99}      \\
\hline

\end{tabular}
 \label{src}
\end{center}
\end{table*}

We plotted the histogram of the pixel values of the dirty maps within the primary beam (FWHM$\sim$37$''$) and identified significant excess positive signals. We found that detections at any positions with S/N $>$ {3.8}\,$\sigma$ can be claimed to be robust, {where the noise used to generate the S/N maps is the r.m.s. of the dirty maps (given in Column~{6} of Table \ref{sma})}. As an example, in Figure \ref{histo}, we show the histogram of the S/N map of Chen-3, along with the ideal Gaussian distribution having r.m.s. of 1 (green curves). We generated the histogram using the the r.m.s. value of the dirty map ($\sigma$; lower panel), and apparently sources with S/N $>$ {3.8}\,$\sigma$ are robust detections in our maps.

{After identifying the detections, }we then performed a deconvolution on the dirty map using the CLEAN routine on the identified source. We CLEANed the area around the detection within an approximately $5''$ box centered at the peak of the source to around 1.5\,${\sigma}$. Note that the size of the box was chosen to be large enough to include all real emission but not too big to enclose spurious noise spikes, which would contaminate the real signals in the CLEAN process. Note also that the resulting source fluxes are not sensitive to the depth to which we chose to clean. We then repeated the process on the residual maps with the identified sources removed to look for sources that may appear after CLEANing. We iterated this process until there were no excess signals{, meaning the S/N distribution agrees with that of pure noise.} 

We primary beam-corrected the fluxes of the CLEANed sources by dividing the CLEANed fluxes by the off-axis gain. We used the IMFIT routine to fit the primary beam-corrected signals to a clean beam---an elliptical Gaussian fitted to the central lobe of the dirty beam---to obtain the fluxes and positions of the detected sources. {The errors from IMFIT correlate with the noise of the CLEANed maps and with the signal-to-noise ratio of the detections. For each detection, we determined both {point-source sensitivities from the dirty maps}, and the errors from IMFIT; however, we adopt the errors from IMFIT for our subsequent detailed analysis of each individual source, since they are more realistic.} Given the small size of the synthesized beam along with the high S/N of our detections, the typical positional uncertainties of our sources are very small (0$\farcs$2--0$\farcs$3 in both R.A. and Decl.), which is critical to estimating the amplifications of strongly lensed sources \citep{Chen:2011p11605}. 

\subsection{SCUBA-2 Observations}
Recently, we have conducted single-dish 850\,$\mu$m surveys on all three massive lensing cluster fields using the novel camera SCUBA-2 \citep{Holland:2013lr} mounted on the JCMT. SCUBA-2, the successor to SCUBA, has an order of magnitude faster mapping speed thanks to a two orders of magnitude increase in the total number of bolometric {detectors}. Its greatly enhanced imaging capability at both 850\,$\mu$m and 450\,$\mu$m makes it possible to take deep submillimeter images with excellent efficiency. Even with only 10--15 hours of observing time, the SCUBA-2 maps of all three fields reach similar depths to the SCUBA maps but with a factor of 20 more sky coverage. The SCUBA-2 observations can therefore provide independent measurements of our sample sources. The details of the SCUBA-2 observations and data reduction can be found in \citet{Chen:2013fk, Chen:2013gq}.

\subsection{{\it Hubble Space Telescope} and {\it Spitzer Space Telescope} Observations}
We also made use of archival data from the {\em Hubble Space Telescope\/} (hereafter, {\it HST}) and the {\em Spitzer Space Telescope\/} (hereafter, {\it Spitzer}). The A370 Advanced Camera for Surveys (ACS) images were taken using the F475W, F625W, and F814W filters with around $6.8$~ks, 2.0\,ks, and 3.8\,ks of exposure (PID: 11507). The ACS F850LP filter was used to take images of A2390 with $\sim$6.4\,ks of exposure (PID: 10504), and the ACS F814W filter was used to take images of A1689 with $\sim10.7$\,ks of exposure (PID: 11710). {We measured the aperture photometry of our SMA detections on the NIR images taken by the Wide Field Camera 3 (WFC3) using the F125W filter. The WFC3 F125W images are, however, only available on A2390 (PI: J. Rigby; PID: 11678) and A1689 (PI: H. Ford; PID: 11802). We estimated the sensitivities of the F125W images using Gaussian fits to the fluxes measured using 1$''$ radius aperture at random source-free positions, yielding 1\,$\sigma$ values of 26.3 and 26.0 AB magnitude for A2390 and A1690, respectively.} 

We retrieved the Super Mosaics, the enhanced data products generated by the Spitzer Science Center (SSC), of all three fields from the {\it Spitzer} archive. The Super Mosaics are produced by combining individual {\it Spitzer} observations and provide the deepest {\it Spitzer} images possible from the archive. We made use of the Super Mosaics from the Infrared Array Camera (IRAC) at 3.6, 4.5, 5.8, and 8\,$\mu$m, and the Multiband Imaging Photometer of Spitzer (MIPS) at 24\,$\mu$m. We list the median exposure times for each field in Table \ref{spitzer}. Throughout this paper, we measure the source fluxes and the upper limits using circular apertures with diameters of 4$\farcs$8, 4$\farcs$8, 6$\farcs$0, 6$\farcs$0, and 18$\farcs$0 at 3.6, 4.5, 5.8, 8.0, and 24\,$\mu$m, which are roughly three times the FWHM of the PSFs \citep{Fazio:2004lr, Rieke:2004lr}. We estimated the sensitivities using Gaussian fits to the fluxes measured at random source-free positions. We give the 1\,$\sigma$ limits in Table \ref{spitzer}, all in $\mu$Jy.

\subsection{Very Large Array Observations}

Deep VLA data were taken at 1.4\,GHz using the A configuration (A2390) and both the A and B configuration (A370). The A2390 (A370) image reaches a 1\,$\sigma$ noise level of 5.6 (5.7)\,$\mu$Jy/beam around the cluster regions with a synthesized beam of $\sim$1$\farcs$4 (1$\farcs$7). The details of the radio images can be found in \citet{Wold:2012fk}.

We also make use of the archival VLA image at 1.4\,GHz of A1689 (PID: AB879). This image is much shallower than the A2390 and A370 images (1\,$\sigma$ $\sim$0.15\,mJy/beam) with a larger synthesized beam of $\sim$6$''$.

\subsection{LENSTOOL}

Throughout this paper, we use LENSTOOL \citep{Kneib:1996p3751}{\, which models three-dimensional mass distributions within the cluster,} to de-lense the sources on the image plane to the source plane in order to calculate the magnification factors due to lensing. We adopted the latest {LENSTOOL} mass models of A370 ($z=0.38$), A1689 ($z=0.18$), and A2390 ($z=0.23$) from \citet{Richard:2010fk}, \citet{Limousin:2007fj}, and \citet{Richard:2010gk}. {The total number of mass components adopted in the lensing models of A370, A1689, and A2390 are 60, 192 and 50, respectively. Note that there are many other mass models available, especially for A370 and A1689, both in the literature (e.g., \citealt{Coe:2010aa}) and on the publicly available Hubble Frontier Fields website\footnote[7]{http://archive.stsci.edu/prepds/frontier/lensmodels/}. 

We also note that lensing magnifications ($\mu$) can be sensitive to the cluster mass models, in particular in strong lensing regions with $\mu >$ 10, where the values can be scattered by a factor up to 40\% due to the degeneracy of different mass model fits \citep{Coe:2010aa}. However, as we show in the discussion section, our main conclusion is not sensitive to this uncertainty.} 
 
\begin{figure}
 \begin{center}
    \leavevmode
      \includegraphics[scale=1]{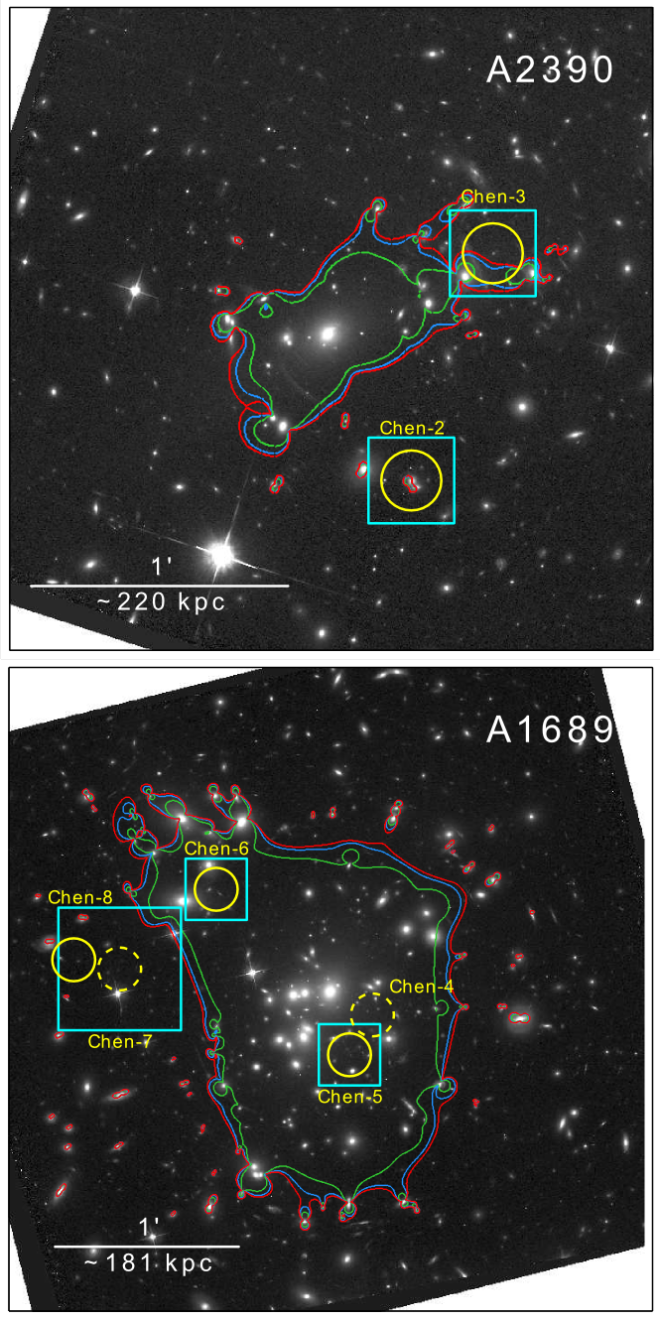}
       \caption{{ 
       Two optical images centered on the brightest cluster galaxy (BCG) of the massive galaxy clusters A2390 and A1689. In each panel, the critical lines at $z=2,4,6$ are drawn as green, blue and red curves, respectively. The SMA observed SCUBA sources are marked in yellow. The SCUBA sources that are SMA detected (undetected) are denoted by solid (dashed) circles. The cyan squares outline the regions where zoom-in images will be presented later in the paper. 
The radius of the circles is 7$\farcs$0, which matches the size of the SCUBA beam FWHM. 
}}
    \label{overview}
  \end{center}
\end{figure}

\begin{figure*}
 \begin{center}
    \leavevmode
      \includegraphics[scale=0.46]{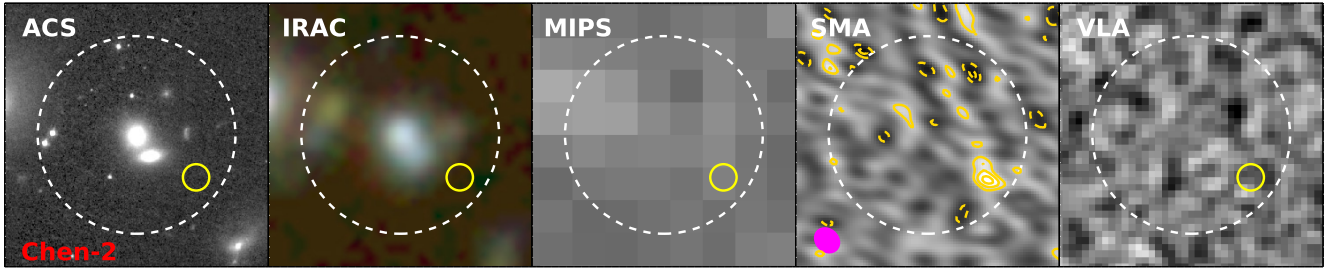}
       \caption{ 
       Postage stamp images centered on the SMA phase center position of {Chen-2}, which is the original SCUBA centroid from \citet{Cowie:2002p2075}. From left to right: 20$'' \times 20''$ gray scale ACS f850lp, IRAC false color (r-g-b) 8.0-5.8-3.6\,$\mu$m, and gray scale MIPS 24~$\mu$m, SMA 870~$\mu$m, and VLA~1.4 GHz images. In each panel {(except SMA)}, we denote the SMA detection by a 1$''$ radius yellow circle. {In the SMA panel, the contours are (-3,-2,2,3,4) $\times \sigma$, and the synthesized beam is presented in the bottom-left corner (magenta).} North is up and East is to the left. The white dashed circle in each panel shows the 7\farcs5 radius SCUBA beam. 
       }
    \label{a23903}
  \end{center}
\end{figure*}

\section{Results}
We obtained SMA detections of 5 of the 8 observed SCUBA sources. We summarize our results in Table \ref{src}. Two of the three SMA undetected SCUBA sources, {Chen-1} and {Chen-4}, are also not detected in the latest SCUBA-2 observations. Both were detected at less than the 4\,$\sigma$ level in the SCUBA data, and we therefore conclude that they are likely to be spurious. {The other SMA undetected source, Chen-7, which is detected in both the SCUBA and SCUBA-2 observations, could be composed of multiple faint sources that are below our current SMA detection limit (more detail in Section 3.3). In Figure \ref{overview}, we show the optical images of A2390 and A1689 with the targeted SCUBA sources marked in yellow. We denote which of the SCUBA sources were SMA detected or undetected by using solid or dashed circles, respectively. We outline the regions with cyan squares where we will be presenting zoom-in images later.} Below we describe each detection in detail.

\subsection{{Chen-2}}
In \citet{Chen:2011p11605}, we reported that {Chen-2 (named {A2390-3} in \citealt{Chen:2011p11605})} had resolved into two distinct sources, {Chen-2a} and {Chen-2b}, located close to one another with a projected angular distance of a few arcseconds. After reanalyzing the maps by CLEANing one source at a time, as opposed to CLEANing both sources together, we found that the significance of the {Chen-2b} detection dropped below 3\,$\sigma$. It is likely that {Chen-2b} is a noise peak boosted by the sidelobes of the real detection, {Chen-2a}. We therefore revise our results for {Chen-2} and present it as a single source detection in Table \ref{src}. 

We show postage stamp images of {Chen-2} in Figure \ref{a23903}, centered at the original SCUBA position from \citet{Cowie:2002p2075}, with the SMA detection denoted by a yellow circle. As we discussed in \citet{Chen:2011p11605}, the fact that {Chen-2} is not detected in any other waveband, and, in particular, in the radio, indicates that this source could be a high-redshift faint SMG.
Radio data are usually an excellent tracer for SMGs, thanks to the well-known empirical correlation between non-thermal radio emission and thermal dust emission among star-forming galaxies \citep{Condon:1992p6652}. Moreover, while the submillimeter flux remains almost invariant over the redshift range $z\sim1-8$ due to a negative $K$--correction (Blain et al.2002), the radio flux drops at high redshifts due to a positive {\it K}--correction. Thus, millimetric redshifts can be estimated from the radio to submillimeter flux ratios with the assumption of a local template spectral energy distribution (SED)
(e.g., \citealt{Carilli:1999p6658, Barger:2000p2144, Ivison:2002uq, Barger:2012lr, Chen:2013fk}). {
While dust properties, such as temperature and emissivity, could in principle cause uncertainties in the millimetric redshifts, it is not known how much and in what way. Reassuringly, \citet{Barger:2012lr} found that their milimetric redshifts agreed well with the spectroscopic redshifts for their SMA observed SMGs, in particular at $z > 3$, where the uncertainties are mainly due to the errors in the flux measurements. 
}

We estimate the millimetric redshift of {Chen-2} by assuming the SED of the local starbursting galaxy Arp~220 ($T_d$ = 47\,K, $\beta$=1). Because {Chen-2} is not detected in the deep VLA map, we use the 3\,$\sigma$ radio flux upper limit of 16.8\,$\mu$Jy.  Based on Equations~2 and 4 in \citet{Barger:2000p2144}, the millimetric redshift of {Chen-2} is then $>$ {3}. Thanks to the two cluster members close to the center of the SCUBA beam, {Chen-2} is strongly amplified by a factor of 4.8$^{+0.5}_{-0.25}$, where the errors represent the uncertainties on the positions ($\pm$\,0$\farcs$2) and redshifts (${3}< z < 6$). The intrinsic (de-lensed) 870\,$\mu$m flux is 0.8$\pm${0.25}\,mJy. Assuming an Arp~220 SED, the total IR luminosity (8--1000\,$\mu$m) of {Chen-2} would be {6.2}$\times10^{11}-{1.1}\times10^{12}~L_\odot$.

\subsection{{Chen-3}}

\begin{figure*}
 \begin{center}
    \leavevmode
      \includegraphics[scale=0.46]{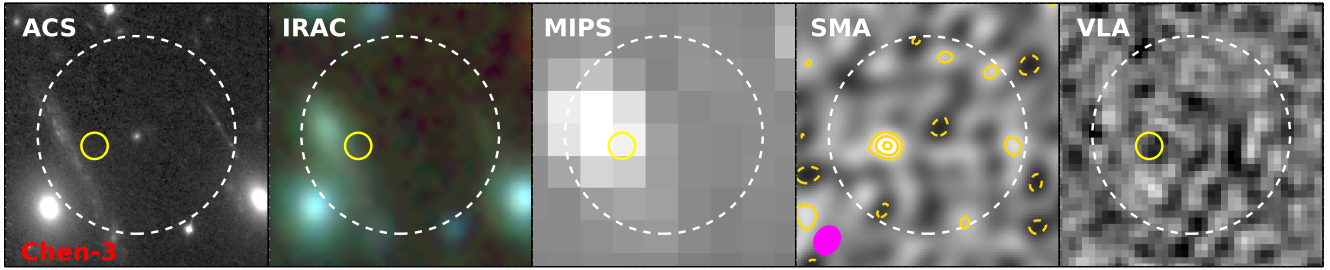}
       \caption{Same as Figure \ref{a23903}, but for {Chen-3}.}
    \label{a23905}
  \end{center}
\end{figure*}

{Chen-3} was first detected by \citet{Cowie:2002p2075} {(source \#14 in \citealt{Cowie:2002p2075})} with an 850~$\mu$m flux of 2.64~mJy. Using ISOCAM, \citet{Metcalfe:2003by} also reported detections at 7 and 15~$\mu$m of the arc structure enclosed by the SCUBA beam. The forbidden cooling line [O II] (3727~\AA) was identified toward this arc by \citet{Pello:1991qy}, indicating a star-forming galaxy at $z=0.913$ and a very likely counterpart for {Chen-3}. However, our SMA observations reveal a different story.

In Figure \ref{a23905}, we show postage stamp images of {Chen-3} with the SMA detection 
denoted by a yellow circle. The only candidate counterpart to {Chen-3} is the IRAC and MIPS detections north-east of the SMA position and located within the SCUBA beam (white dashed circle). 
However, referring to the ACS image, it is more likely that the IR candidate counterpart traces part of the optical arc instead of the submillimeter signal from {Chen-3}, based on the fact that the morphology of the IR signal is elongated (aligned with the arc), and the positional offset is much greater than the positional error measured from the SMA.

\begin{figure}
 \begin{center}
    \leavevmode
     \includegraphics[scale=0.45]{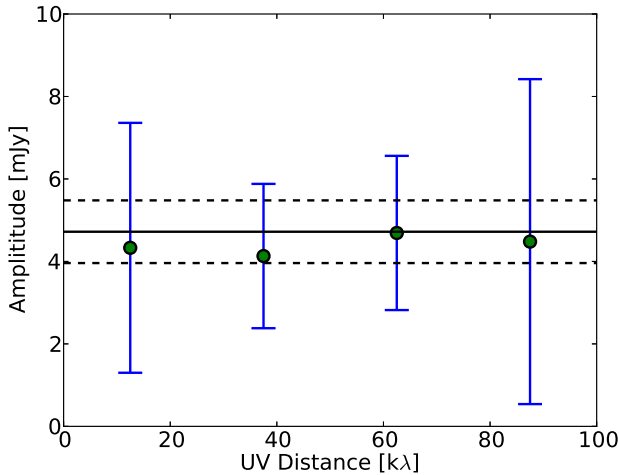}
       \caption{Real amplitude in mJy vs. the {\it uv} distance for {Chen-3}. The black solid line represents the primary beam corrected fluxes obtained using IMFIT on the CLEANed image,
       and the dashed lines show the 1\,$\sigma$ errors.}
    \label{a23905uvamp}
  \end{center}
\end{figure}

Again, surprisingly, we do not see any counterpart to this faint SMG at other wavelengths. A non-detection in the deep radio map (3\,$\sigma$ of 16.8\,$\mu$Jy) gives a millimetric redshift of {$z >$ 3.5}. Due to its close proximity to the critical lines {(Figure \ref{overview})}, {Chen-3} is highly amplified by a factor of at least 45, considering its positional ($\pm$\,0$\farcs$2) and redshift (${3.5}<z<6$) uncertainties. We therefore adopt 45 as the nominal amplification of {Chen-3}. {
We caution that, as stated in Section 2.5, because Chen-3 is located at a position very close to the critical lines, the scatter of the magnifications could be large. However, given its position close to the cluster center and its alignment with the orientation of the cluster mass distribution, the minimum amplification is likely to be larger than 2, which makes Chen-3 likely to be a faint SMG with an intrinsic flux $<$ 2\,mJy. 
}

Interestingly, although {Chen-3} is expected to be highly amplified and stretched, the source itself appears to be point-source-like. In Figure~\ref{a23905uvamp}, we show the flux versus the {\it uv} distance for {Chen-3}. A flat trend indicates that the source is unresolved. It could be that {Chen-3} is extremely compact and that the current resolution is not sufficient to resolve the source.

By adopting an amplification factor $>45$, {Chen-3} has an intrinsic 870\,$\mu$m flux of $<$0.12\,mJy. Again assuming an Arp~220 SED and a redshift of ${3.5}<z<6$, {Chen-3} has a Milky Way like total IR luminosity of $< 10^{11}\,L_\odot$. {If its low intrinsic IR luminosity were to be confirmed, Chen-3 would be} a source with a relatively modest luminosity that is completely hidden from deep optical/NIR/radio observations. Sources like {Chen-3} would be completely missed in current optical/NIR calculations of the cosmic star formation history; however, given the amount of light in the faint SMG population, they would contribute comparable amounts of star formation.

\subsection{{Chen-4, 5, 6, 7, 8}}

\begin{figure*}
 \begin{center}
    \leavevmode
      \includegraphics[scale=1.18]{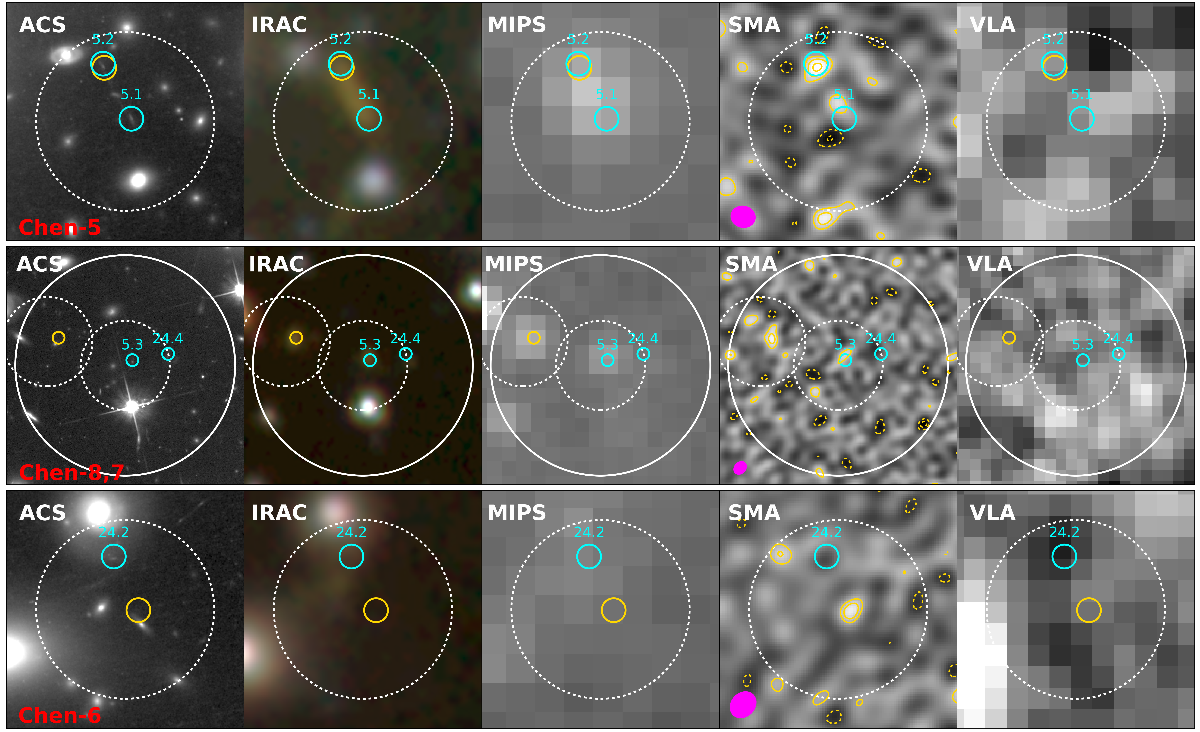}
       \caption{Postage stamp images of (top row) {Chen-5}, (middle row) {Chen-7} and {Chen-8}, and (bottom row) {Chen-6}. The white dashed circles show $7\farcs5$ radius SCUBA beams. The middle row contains two SCUBA sources: {Chen-7} (center) and {Chen-8} (left). Here, the size of each panel is adjusted to $40'' \times40''$ to enclose both sources. For the other two rows, the size of each panel is $20''\times20''$. The white large solid circle shows the size of the SMA primary beam with 50\% sensitivity relative to the phase center. The yellow circles mark the SMA detections. The cyan circles mark the positions of the multiple images from background sources 5 and 24, as labeled. The optical image is from the ACS filter F814W.}
    \label{multismm}
  \end{center}
\end{figure*}

The SCUBA discoveries of {Chen-4, 5, 6, 7, and 8} were first reported by \citet{Knudsen:2008p3824}, where they were called SMM J131128.6-012036, SMM J131129.1--012049, SMM J131132.0--011955, SMM J131134.1--012021, and SMM J131135.1--012018, respectively. {Chen-4/Chen-5 and Chen-7/Chen-8 are two source pairs that} are located close to one another. 
The fluxes of {Chen-5},  {Chen-6}, and {Chen-7} were suspected by Knudsen et al.to be contributed by a combination of lensed multiple images of two background sources at $z\sim2.6$, denoted as source \#5 and source \#24 in the study of the A1689 mass model \citep{Limousin:2007fj}. We show the positions of the multiple images in Figure~\ref{multismm} as cyan circles with the identifications (background source number dot multiple image number) marked. Only one SMA detection ({Chen-5}) is aligned with a lensed multiple image (5.2). 

\begin{figure}
 \begin{center}
    \leavevmode
     \includegraphics[scale=0.43]{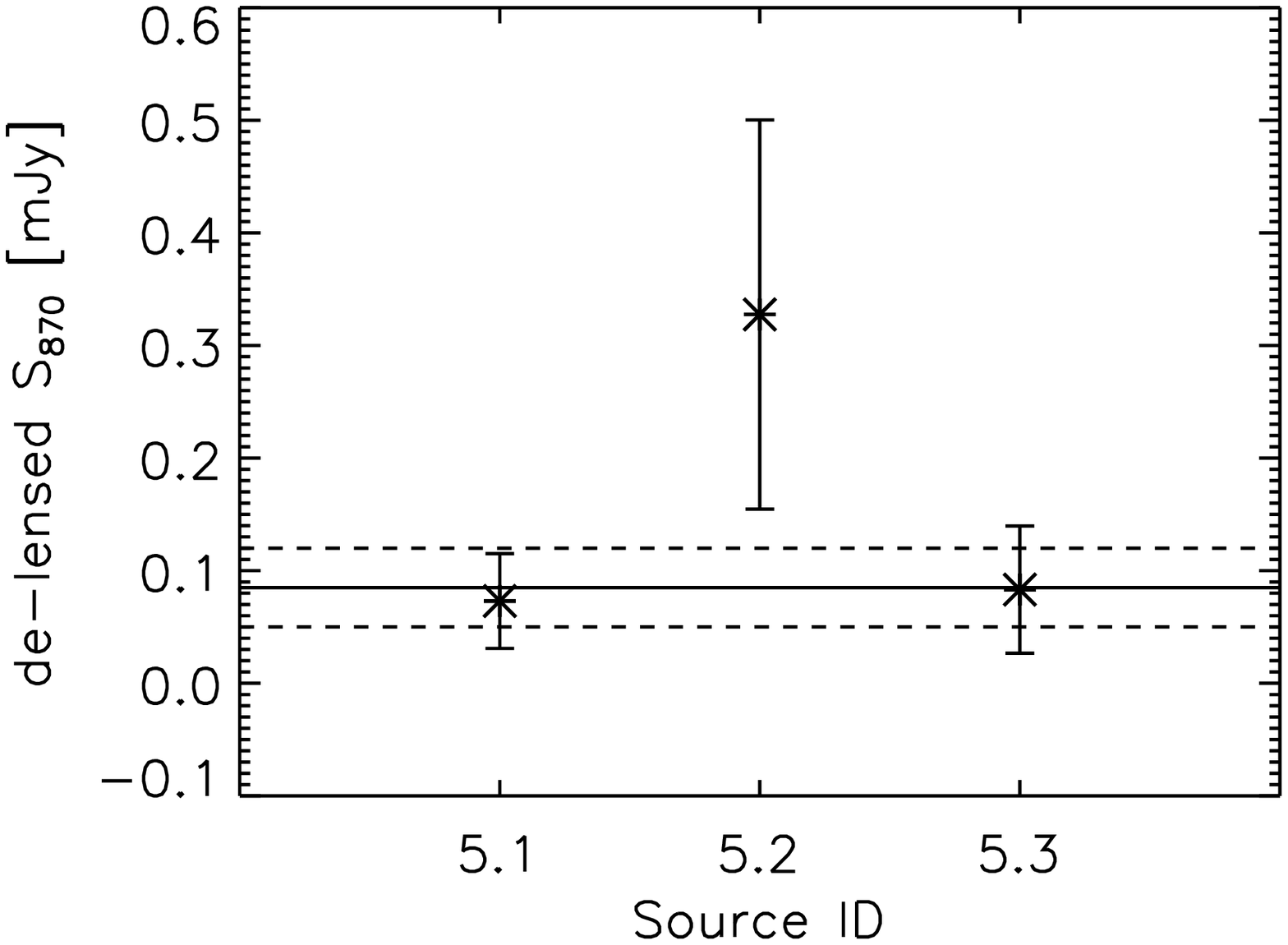}
       \caption{De-lensed 870\,$\mu$m fluxes with {1\,$\sigma$} errors obtained by de-lensing the measured submillimeter fluxes at the nominal positions of the multiple optical images of the background source \#5. The best fit de-lensed flux is shown with the black line, while the errors are plotted with dashed lines. {The magnifications adopted in this calculation are 31.9, 12.8 and 11.6 for images 5.1, 5.2, and 5.3, respectively.}}
    \label{deflxs5}
  \end{center}
\end{figure}

Based on the {mass model of \citet{Limousin:2007fj}}, the magnification of image 5.1 is $\sim2.5$ times higher than that of image 5.2. If the background source is indeed the origin of the submillimeter emission, then we should detect a submillimeter flux toward 5.1 that is stronger than what we observe. Our lack of an SMA detection of image 5.1 could imply that the submillimeter emission is not related to the background source \#5. However, we do observe faint submillimeter emission at the positions of images 5.1 and 5.3, so the lack of significant detections could also be caused by noise. 

{ 
Alternatively, the lensing magnifications of images 5.1 and 5.2 are subject to large uncertainties due to degeneracies of the mass models, so the lack of a detection on image 5.1 could also be caused by the lensing uncertainties. With more data on the background galaxies and multiple images in A1689, \citet{Coe:2010aa} also computed magnification estimates on images 5.1 and 5.2. Although their final adopted values are similar to those obtained using the model of \citet{Limousin:2007fj} in the sense that image 5.1 is amplified by a larger factor than that of image 5.2, \citet{Coe:2010aa} found that among their ensemble of models, the mean magnification of image 5.2 (5b in \citealt{Coe:2010aa}) was larger than that of image 5.1 (5a in \citealt{Coe:2010aa}), with a scatter of $\sim$50\% on both values. 
}

We examine the possibility {that source \#5 is the counterpart of the submillimeter emission} by measuring the submillimeter fluxes at the nominal positions of the multiple optical images of source \#5 and comparing their de-lensed fluxes. If the submillimeter emission comes from the same background source, then the de-lensed fluxes should be the same within the errors. We include both the uncertainties on the flux measurements and the uncertainties on the magnification estimates {($\sim$50\%; \citealt{Coe:2010aa})} in our error calculations. We show the results in Figure \ref{deflxs5}: the de-lensed 870\,$\mu$m fluxes of the multiple optical images are indeed the same within errors. We thus conclude that the lensed images 5.1 and 5.2 of background source \#5 are likely to be the counterparts of submillimeter source {Chen-5} and that the lensed image 5.3 contributes to {Chen-7}. With the constraints from the multiple measurements, we find the best fit intrinsic 870\,$\mu$m flux of {Chen-5} is {0.085$\pm$0.035\,mJy} (black line in Figure \ref{deflxs5} with errors in dashed lines). At $z=2.6$, the total IR luminosity of {Chen-5} is $\sim$ {8.5$^{+3.5}_{-3.5}\times10^{10}$}\,L$_\odot$ assuming an Arp\,220 SED.

\begin{figure}
 \begin{center}
    \leavevmode
      \includegraphics[scale=0.35]{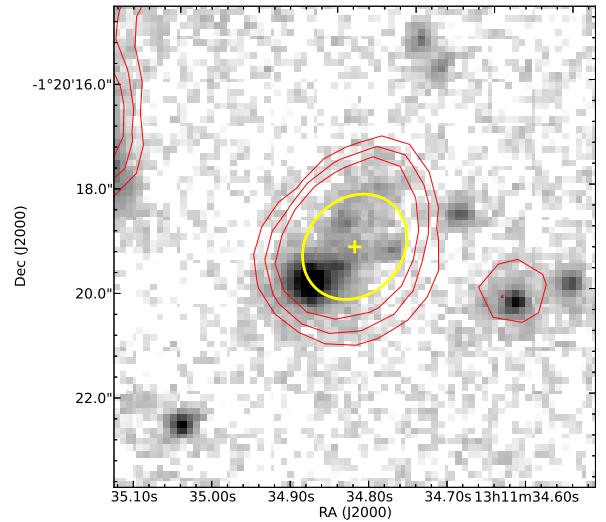}
       \caption{Zoomed-in inverse grey scale F140W image of {Chen-8} with red contours representing the 3.6\,$\mu$m IRAC emission. The shape of the SMA synthesized beam is shown with the yellow circle.  The yellow cross marks the SMA position with the positional errors shown in its length. The angular width of this map is 10$''$.}
    \label{smm131135zoom}
  \end{center}
\end{figure}

The neighboring pair system {Chen-7} and {Chen-8} is detected in both the SCUBA and SCUBA-2 observations with consistent flux measurements (Table \ref{src}). Interestingly, we only detect {Chen-8} with the SMA. The fact that {Chen-7} is detected in both the SCUBA and SCUBA-2 observations but not in the SMA observations suggests that it may be composed of multiple faint sources that are below our current SMA detection limit. This would include image 5.3 which, based on the lensing discussion above, would have an observed 850~$\mu$m flux of $\sim$1\,mJy.

{Chen-8} is detected by all three independent observations (SCUBA, SCUBA-2, and SMA). We find counterparts in the IRAC and MIPS images. We also find low surface brightness extended emission in the NIR images obtained from {\it HST} WFC3 toward the SMA position. We show the zoomed-in inverse grey scale F140W image in Figure~\ref{smm131135zoom}. We mark the SMA synthesized beam shape with a yellow circle. While the IR emission seen in the IRAC and MIPS images could be contributed by both the point source to the south-east and the extended structure, it is perhaps more likely that most of the submillimeter flux comes from the extended structure, given the SMA position and its optically faint nature. Due to the fact that the IR fluxes are contributed by both the bright point source and the extended structure, it is unrealistic to estimate the photometric redshift of {Chen-8} using the IR flux measurements. However, based on the fact that it is not detected in the radio images with 3\,$\sigma$ = 0.45\,mJy, we can put a lower limit on its redshift as $z>0.5$. {Chen-8} is also strongly amplified by a factor of 3--9, depending on its redshift from $z=$0.5$-$6. The intrinsic fluxes of the IR counterparts are 22.1--23.3, 21.6--22.8, 21.4--22.6, 21.4--22.6, and 19.4--20.6 in AB magnitudes or 1.8--5.4, 2.7--8.0, 3.2--9.6, 3.3--9.9 and 20--60\,$\mu$Jy at 3.6, 4.5, 5.8, 8.0 and 24\,$\mu$m for the same redshift range. By adopting the nominal amplification of 6, the intrinsic 870\,$\mu$m flux would be 0.65$^{+0.86}_{-0.28}$\,mJy, and assuming an Arp\,220 SED the total IR luminosity would be $\sim$4--6$\times 10^{11}$\,L$_\odot$.

{Chen-6 was first discovered by \citet{Knudsen:2008p3824} and the detection is confirmed by both our SCUBA-2 and SMA observations with consistent flux measurements.} We mark {the} SMA sources in yellow in the bottom row of Figure~\ref{multismm}. Interestingly, it is not located close to the lensed multiple image of the background source \#24, which was suspected by \citet{Knudsen:2008p3824} to be the main contributor of the observed submillimeter flux. Moreover, as we see in most of our SMA detections, there is no obvious counterpart found in observations at any other wavelength. There is an optical candidate counterpart lying on the edge of the beam; however, it is unlikely to be the true counterpart, since it is optically bright but IR faint, and the positional offset is significant.

{Chen-6} lies close to two massive cluster members, seen in the figure lying just outside the white dashed circle, and is strongly magnified. Again, the lack of a radio detection for {Chen-6} implies that {it is a} background source behind the cluster ($z>0.5$). Although without the redshift information it is hard to determine the amplification factors for sources that are strongly magnified, we can determine lower limits, as we did for {Chen-3}. The lower limit is 10 over the redshift range $z = 0.5$--6 for {Chen-6}, which gives the intrinsic 870\,$\mu$m fluxes of $<$0.34\,mJy. Assuming an Arp\,220 SED, the total IR luminosity of {Chen-6} would be $<$4$\times 10^{11}$\,L$_\odot$.

\begin{table*}
\begin{center}
 \caption{The properties of the SMA detected sources}
\begin{tabular}{llllrc}
\hline
\hline
 ID              &  Magnifications   &  S$_{\mathrm{870,intrinsic}}$  &  S$_{\mathrm{F125W,intrinsic}}$ &    $z^b$  & log($L_{8-1000\,\mu m})$\\
                 &                  &  (mJy)             &(mag)   &       &   \\
\hline
 {Chen-2}         &  4.8 (4.7--5.3)  &  0.83 (0.64--0.97)    &  $>$\,26.8 & $>$\,{3.0} & 11.8--12.0\\
 {Chen-3}         &  45 ($>$\,45)      &  0.12 ($<$\,0.12)        & $>$\,28.2     & $>$\,{3.5} & $< 11.1$ \\
 {Chen-5}   &  19 (16--22)     &  {0.09 (0.05--0.12)}$^a$    & 26.6 (26.5--26.8) &2.600 & {10.7--11.1}$^a$\\
 {Chen-6}  &  10 ($>$\,10)     &  0.34 ($<$\,0.34)    & $>$\,25.6 &$ >$\,0.5 & $< 11.6$\\
 {Chen-8}   &  6 (3--9)        &  0.65 (0.37--1.51)    & $\cdots$  &$>$\,0.5 & 11.6--11.8\\
\hline
&&& \\
 \multicolumn{6}{l}{$^{a}$ Mean values based on the measurements on all three multiple images of source \#5.} \\
 \multicolumn{6}{l}{$^b$ Lower limits are millimetric redshifts estimated using the ratio between the radio and the} \\
 \multicolumn{6}{l}{ submillimeter fluxes.} \\
\end{tabular}
 \label{delen}
\end{center}
\end{table*}

\section{Discussion}

Because the positions of the single-dish detected SMGs are poorly determined, accurately finding the true SMG counterparts is critical for understanding their characteristics. Since SMGs are dusty and their emission appears to be dominated by star formation (e.g., \citealt{Alexander:2005p6453}), sources detected in {\it Spitzer} MIR and VLA images with better determined locations are often used to cross-identify the SMG counterparts (e.g., \citealt{Ivison:2007fj,Biggs:2011uq}). However, both MIR and radio fluxes drop significantly at high redshifts, while the 850\,$\mu$m fluxes remain almost invariant over the redshift range $z = 1$ -- 8 due to the negative {\it K}-correction \citep{Blain:2002p8120}. Many high-redshift, dusty sources are inevitably missed in flux-limited observations at MIR and radio wavelengths.

Recently, observations using ALMA with arcsecond level spatial resolution have successfully pinpointed the location of a flux-limited sample of $\sim$100 870\,$\mu$m SMGs selected by the LABOCA survey in the ECDF-S field \citep{Hodge:2013lr}. Armed with accurate positions, Hodge et al.tested the robustness of the counterpart identifications made with MIR and radio data by \citet{Biggs:2011uq}. They found that only 45 out of their 99 robustly detected sources (ALESS MAIN) had robust MIR/radio counterparts; the recovery rate increased to $\sim$55\% if they included tentative MIR/radio identifications. 

Interestingly, if we separate the Hodge et al.sample into flux bins, then the fraction of SMGs with robust MIR/radio counterparts (black circles in Figure \ref{detrat}) dramatically decreases for bins fainter than 3\,mJy. Our sample extends even fainter, and we obtain a similarly low fraction (2 out of our 6 SMA sources have MIR/radio counterparts; blue circle in Figure \ref{detrat}). Note that the depth of the MIR/radio images is key to the results shown in Figure \ref{detrat}, as we expect that more sources could be recovered with deeper images. Indeed, \citet{Barger:2012lr} recently showed that all their bright SMGs with 860\,$\mu$m fluxes above 3\,mJy are recovered by ultradeep 1.4\,GHz images (1$\sigma$$\sim$2.5\,$\mu$Jy). 

With the strong gravitational lensing (amplification $>$ 5), the {\it Spitzer} images in our source fields (1\,$\sigma$; $<$0.08, $<$0.08, $<$0.3, $<$0.5, and $<$8\,$\mu$Jy at 3.6, 4.5, 5.8, 8.0, and 24\,$\mu$m) are deeper than ECDF-S and almost as deep as in GOODS-S, which is one of the deepest {\it Spitzer} observed fields \citep{Magnelli:2009fk, Damen:2011lr}. 

Similarly, with the strong lensing, our A2390 radio images reach 1\,$\sigma$ $<$ 1\,$\mu$Jy at our source positions, by far the deepest 1.4\,GHz depth. Even in A1689, where the radio data are shallower, with the lensing, the sensitivity reaches 1\,$\sigma$$\sim$10\,$\mu$Jy, similar to the depth of the ECDF-S \citep{Miller:2008qy}.

\begin{figure}
 \begin{center}
    \leavevmode
      \includegraphics[scale=0.85]{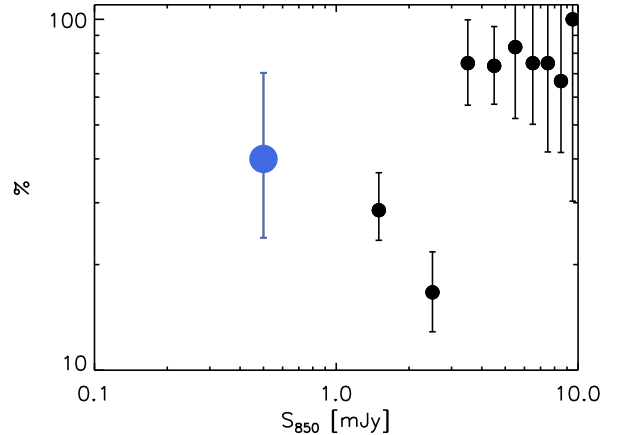}
       \caption{The percentage of SMGs in each flux bin robustly identified in MIR/radio images with similar depths to the ECDF-S field. The SMGs shown in this figure were all observed with arcsecond resolution submillimeter interferometers. Black circles are from \citet{Hodge:2013lr}, and the blue circles shows our work.}
    \label{detrat}
  \end{center}
\end{figure}

We can also see this effect in the optical/NIR regime. We measured 1$''$ radius aperture F125W magnitudes at the SMA positions using the {\it HST} WFC3 archival images {described in Section 2}. In Table \ref{delen}, we summarize these measurements along with other detailed characteristics of each SMA detection, including the magnifications, de-lensed fluxes, and redshifts. We show the histogram of our F125W magnitudes in dark blue in Figure \ref{opflx}, excluding only {Chen-8}, where there is no F125W imaging. For the cases where there was no detection in F125W, which applies to most of our SMA sources, we calculated the upper limits by taking into account the lensing uncertainties. We corrected all of our measurements to total magnitudes based on the released encircled energy fractions (Table 7.6 in \citealt{Dressel:2012kx}). We de-lensed them based on the adopted magnifications. We estimated any contamination due to foreground emission from cluster members using pixels with distances between 1$\farcs$2 and 1$\farcs$4 from the SMA positions. 

For comparison, we show the histogram of the F125W magnitudes of the SMA-detected, bright SCUBA-2 SMG sample with 860\,$\mu$m fluxes $>$ 3\,mJy given in Barger et al.(2013) {\em (hatched)}, except CDFN1, CDFN2, CDFN3, and CDFN18, where there is no F125W imaging, and GOODS~850-17, where the flux of the source is too contaminated by the neighboring source \citep{Barger:2012lr}. We measured the magnitudes using a 1$''$ radius aperture on the {\it HST} WFC3 archival images obtained for the Cosmic Assembly Near-IR Deep Extragalactic Legacy Survey (CANDELS; \citealt{Grogin:2011fj}) by PI S. Faber (PID: 12443, 12444, 12445), except GOODS~850-1, where we used a 0.5$''$ radius aperture to minimize contamination from the neighboring sources. We corrected all of our measurements to total magnitudes. 

We also show the magnitudes for other faint SMGs in the literature with intrinsic 870\,$\mu$m fluxes $<$ 2\,mJy {\em (light blue)}. These were measured in the $J$-band, except for SMM J163555.5+661300, where the closest passband available was the F110W filter of WFC3 \citep{Knudsen:2010lr}. 

\begin{figure}
 \begin{center}
    \leavevmode
      \includegraphics[scale=0.45]{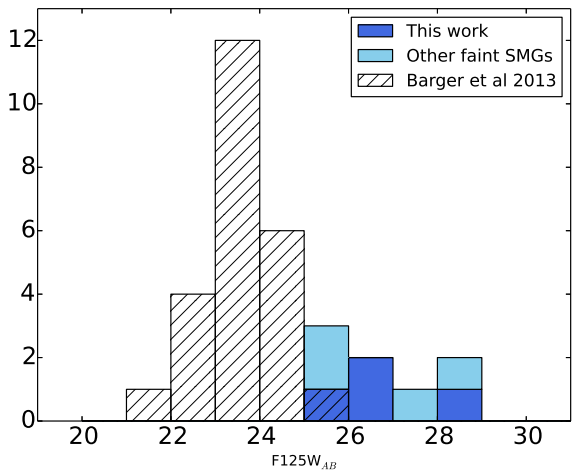}
       \caption{Histograms of the intrinsic F125W magnitudes of our faint SMG sample ({\it dark blue}), other faint SMGs from the literature \citep{Frayer:2003p10289, Kneib:2004uq, Gonzalez:2009fj,Knudsen:2010lr} ({\it light blue}), and the bright SMG sample from Barger et al.(2013) ({\it hatched}).
       The histogram of the faint SMGs is stacked, whereas that of the bright SMG sample is independent from the faint SMGs.}
    \label{opflx}
  \end{center}
\end{figure}

From Figure \ref{opflx}, we see that, in contrast to the bright SMGs, faint SMGs with 870\,$\mu$m fluxes $<$ 2\,mJy are statistically dimmer with F125W AB magnitudes spanning the range 24--29. {We stress that the fact that faint SMGs are statistically dimmer than bright SMGs in the NIR is not affected by the uncertainties of strong lensing. Since the uncertainties of lensing amplifications are affected by many factors, such as the degeneracies of the cluster mass models and the errors in the measurements of the source positions in coordinates and in redshifts, it is difficult to quantify the lensing uncertainties for each individual faint SMG in Table \ref{delen}. However, if we assume all of our sources have only a lensing magnification of 2, which is very conservative given their proximity to the cluster center, the median de-lensed F125W magnitude of the faint SMGs ({25.4} including other faint SMGs from the literature) is still larger than all the bright SMGs. Moreover, four out of our five F125W measurements are upper limits, the median de-lensed F125W magnitude of faint SMGs is likely to be lower than {25.4}. } 

{Our results suggest that} a large fraction of the faint SMGs will be missed in NIR surveys having magnitude limits 22--25 (e.g., \citealt{Quadri:2007yq, Keenan:2010fr}). This is in agreement with the stacking analysis of \citet{Wang:2006p2031}, who stacked the GOODS-N SCUBA data at the positions of sources detected in the NIR with an AB magnitude limit of $\sim$24. They found that after excluding the bright SMGs from the maps, they could account for only about one-quarter of the 850\,$\mu$m EBL (based on the EBL measurement of \citealt{Fixsen:1998p2076}) with their combined $H$-band and 3.6\,$\mu$m sample and that this light was coming from sources at $z < 1.5$. Based on our latest number counts, bright SMGs contribute about another quarter of the 850 \,$\mu$m EBL \citep{Chen:2013fk,Chen:2013gq}, which implies that up to 50\% of the 850 \,$\mu$m EBL is still hidden from deep NIR samples. Similar results were presented by \citet{Serjeant:2008p6307}, who stacked SCUBA data at the positions of sources detected in {\it Spitzer} images at 3.6\,$\mu$m (1.5\,$\mu$Jy, 1\,$\sigma$), 4.5\,$\mu$m (1.5\,$\mu$Jy, 1\,$\sigma$), 5.8\,$\mu$m (3\,$\mu$Jy, 1\,$\sigma$), 8\,$\mu$m (4\,$\mu$Jy, 1\,$\sigma$) and 24\,$\mu$m (9\,$\mu$Jy, 1\,$\sigma$). They also found that after excluding the bright SMGs, they could account for only about one-quarter of the 850\,$\mu$m EBL with their {\it Spitzer} sample, and, similar to \citet{Wang:2006p2031}, that this light was dominated by sources at $z < 1.5$.  

\begin{figure}
 \begin{center}
    \leavevmode
      \includegraphics[scale=0.49]{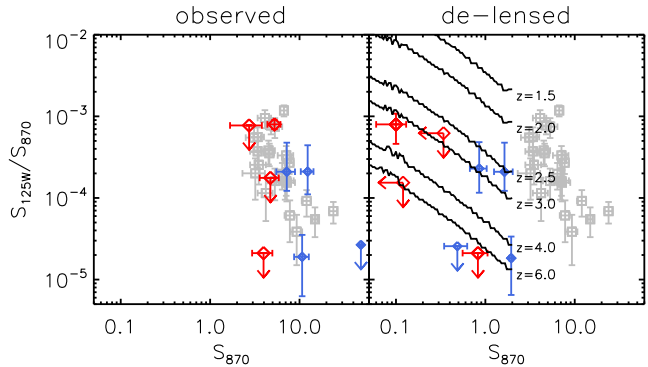}
       \caption{{The F125W-to-870\,$\mu$m flux ratios vs. 870\,$\mu$m fluxes of our SMA detected faint SMGs ({\it red diamonds}), other faint SMGs from the literature that are the same as those used in Figure \ref{opflx} ({\it blue diamonds}), and the bright SMGs ({\it squares}) adopted from Barger et al. (2014). Three out of our four SMA-detected faint SMGs (no F125W data available for Chen-8; Table \ref{delen}) are not detected in the F125W maps; thus, we show their upper limits. In the left panel, we plot the flux ratios against the observed 870\,$\mu$m fluxes, whereas in the right panel, we plot the flux ratios against the de-lensed 870\,$\mu$m fluxes by adopting the information in Table \ref{delen}. In the right panel, we also show the empirical predictions ({\it black curves}) based on the SED templates of \citet{Chary:2001aa} at $z =$ 1.5, 2.0, 2.5, 3.0, 4.0, and 6.0.}}
    \label{ratios}
  \end{center}
\end{figure}

{Consistent with these stacking results, studies of $z <$ 2 (U)LIRGs have shown that there is less dust obscuration (L$_{IR}$/L$_{UV}$) in low-luminosity sources (e.g., \citealt{Chary:2001aa, Le-Floch:2005fk, Reddy:2010aa}), which would suggest that the observed NIR/submm flux ratios should increase as we go from bright SMGs to faint SMGs. This, together with the fact that the lensed faint SMGs are amplified relative to the bright SMGs, leads to the expectation that the observed F125W fluxes of the faint SMGs should be brighter than those of the bright SMGs. However, this is not what we observe. 

In Figure \ref{ratios} (left), we plot the F125W-to-870\,$\mu$m flux ratios versus the observed 870\,$\mu$m fluxes for our SMA detected faint SMGs (diamonds), together with those of the bright SMGs (squares) from \citet{Barger:2014aa}.
In Figure \ref{ratios} (right), we de-lense the faint SMGs, plotting their intrinsic fluxes on the x-axis. We also show the empirical predictions (black curves) based on the SED templates of \citet{Chary:2001aa} for redshifts from 1.5 to 6.0. Figure \ref{ratios} suggests that these faint SMGs must either be at much higher redshifts than what was expected from the stacking analyses, or be extremely dusty.}

All this evidence suggests that many faint SMGs are optically and NIR faint, and that there is a considerable amount of submillimeter light coming from objects which are fainter than the population probed by most of the current NIR, MIR, and even radio surveys. Our results also suggest that there are many low-luminosity, obscured star forming galaxies at high redshift that do not merge into the normal galaxy population and hence would not be included in the optical star formation history.

\section{Summary}
We conducted SMA observations of 8 faint SMGs detected by SCUBA with intrinsic 850\,$\mu$m fluxes $<$ 2\,mJy. We obtained SMA detections of 5 of the SCUBA sources. Based on the latest number counts, these sources contribute $\sim$70\% of the 850\,$\mu$m EBL, and they represent the dominant star-forming galaxy population in the dusty universe. We found that the fraction of faint SMGs with
MIR/radio counterparts is low, {40}$^{+30}_{-16}$\%, compared with bright SMGs, where the majority have counterparts. We also found that the NIR counterparts of faint SMGs are statistically dimmer than those of bright SMGs, suggesting that many faint SMGs must either be at very high redshifts, or be extremely dusty. Our results also suggest that there are many low-luminosity, obscured star forming galaxies at high redshifts that would not be included in measurements of the optical star formation history.

\vspace{5mm}
{\it Acknowledgments}
We thank the referee for comments that improved the manuscript. We gratefully acknowledge support from NSF grants AST-0709356 (C.C.C., L.L.C.), AST-1313309 (L.~L.~C.), and  AST-1313150 (A.~J.~B.), the University of Wisconsin Research Committee with funds granted by the Wisconsin Alumni Research Foundation (A.J.B.), the David and Lucile Packard Foundation (A.J.B.), and the National Science Council of Taiwan grant  
102-2119-M-001-007-MY3 (W.-H.W.). 
We thank SMA support astronomer Glen Petitpas. This research made use of Astropy, a community-developed core Python package for Astronomy \citep{Astropy-Collaboration:2013aa}. The authors wish to recognize and acknowledge the very significant cultural role and reverence that the summit of Mauna Kea has always had within the indigenous Hawaiian community.  We are most fortunate to have the opportunity to conduct observations from this mountain.

\bibliography{bib}

\end{document}